# Improving Embedding Efficiency for Digital Steganography by Exploiting Similarities between Secret and Cover Images

Alan A. Abdulla, Harin Sellahewa, Sabah A. Jassim

*Abstract*. Digital steganography is becoming a common tool for protecting sensitive communications in various applications such as crime/terrorism prevention whereby law enforcing personals need to remotely compare facial images captured at the scene of crime with faces databases of known criminals/suspects; exchanging military maps or surveillance video in hostile environment/situations; privacy preserving in the healthcare systems when storing or exchanging patient's medical images/records; and prevent bank customers' accounts/records from being accessed illegally by unauthorized users. Existing digital steganography schemes for embedding secret images in cover image files tend not to exploit various redundancies in the secret image bit-stream to deal with the various conflicting requirements on embedding capacity, stego-image quality, and undetectibility. This paper is concerned with the development of innovative image procedures and data hiding schemes that exploit, as well as increase, similarities between secret image bit-stream and the cover image LSB plane. This will be achieved in two novel steps involving manipulating both the secret and the cover images, prior to embedding, to achieve higher 0:1 ratio in both the secret image bit-stream and the cover image LSB plane. We exploit the above two steps strategy to use a bit-plane(s) mapping technique, instead of bit-plane(s) replacement to make each cover pixel usable for secret embedding. We shall demonstrate that this strategy produces stego-images that have minimal distortion, high embedding efficiency, reasonably good stego-image quality and robustness against 3 well-known targeted steganalysis tools.

## I. INTRODUCTION

Digital steganography is an alternative information security mechanism to cryptography that is generally concerned with concealing the presence of a secret data/object during mundane communication sessions. While steganographers aim to design efficient and difficult to detect steganography schemes, steganalysers attempt to defeat the goal of steganography by detecting the presence of a hidden message, even if they cannot retrieve it. Digital media files (audio, images and videos) is a rich source of cover files, due to the fact that such files involve large amounts of redundancies, for hiding secrets without having significant impact on the information content, or quality of stego-file. Moreover, images are widely exchanged over the Internet than other digital media and attract little suspicion. This paper focuses on the case where both the secret and cover files are images.

Success criteria in steganography are related to a list of rather competing requirements on: 1) stego-image quality; 2) hiding capacity; 3) secret detectability; and 4) robustness against active attacks. Most existing steganography schemes only attempt to deal with the first three requirements, while robustness is application dependent [1], and most schemes consider the passive warden scenario in which the warden does not interfere with the stego file [2].

Minimizing the number of modified cover pixels post secret embedding improves chances of success. In fact, reducing the ratio of modified pixels to the payload capacity has recently been proposed as an indicator of higher stego-image quality and lower message detectability. Accordingly, the *embedding efficiency* (EE) of a hiding scheme is defined as:

$$\text{EE} = \frac{1}{ratio\ of\ modified\ pixels} \quad (1)$$

The higher EE is the less detectable traces is introduced in the stego-image, and the more robust the scheme is against steganalysis techniques. The primary objective of this paper is to maximize EE values, while maintain high loading capacity. Our main strategy is to improve EE values by exploit existing knowledge of image objects to increase similarities between the secret image bit-streams and least significant bit (LSB) plane of the cover images.

The rest of the paper is organized as follows. Section II reviews the related work. Section III presents the proposed methods for secret image manipulation as well as a new pixel value decomposition scheme followed by explaining a mapping based embedding technique, and embedding and extracting procedure of our proposed steganography schemes. The experimental results are presented and analyzed in Section IV, followed by the conclusion in Section V.

Alan A. Abdulla is presently a lecturer/researcher in computer science at the Department of Information Technology, College of Commerce, University of Sulaimani in Kurdistan Regional Government (KRG) in Iraq, (e-mail:alananwer@yahoo.com). The majority of the work presented in this paper has been done during the PhD research stay at The University of Buckingham in UK.

Harin Sellahewa is a senior lecturer in computer science and a head of the Applied Computing Department, University of Buckingham, Buckingham MK18 1EG, U.K, (e-mail:harin.sellahewa@buckingham.ac.uk).

Sabah A. Jassim is a Professor at The Department of Applied Computing, University of Buckingham, Buckingham MK18 1EG, U.K, (e-mail: sabah.jassim@buckingham.ac.uk).

## II. BACKGROUND AND RELATED WORK

Image-based steganography schemes are classified in terms of cover image domain into spatial and transformed/frequency schemes. In the latter case, secret bits are used to manipulate the transformed cover image coefficients, and the 3 most commonly used transforms are: the Discrete Fourier Transform DFT [3], the Discrete Wavelet Transform DWT [4], and the Discrete Cosine Transform DCT [5]. In general, spatial domain schemes replace/substitute the cover image least significant bit-plane, or other bit-planes, with secret bits. Examples include variants of the least significant bit replacement LSBR scheme. These schemes are easy to implement, have relatively high stego-image quality and payload capacity, and their performance is mostly tested by embedding random secret bit-streams. Relatively minor changes are made to cover pixel intensity and embedding increases/decreases even/odd cover pixel values either by one or leaves it unchanged. However, such schemes distort the statistical distribution in the pairs of pixel values (0, 1); (2, 3); . . . (254, 255), which is known as the *asymmetry* problem [6]. Steganalysis schemes attempt to exploit the fact that any embedding scheme will result in some kinds of local random distortions, albeit difficult to detect by the naked eye, or may violate in a small, but computable, way statistical/correlation models that are known/expected to hold among the different spatial/gray-level components of cover images. For example, the asymmetry problem mentioned above can be exploited to detect the existence of a hidden message using certain steganalysis techniques, even at a low embedding rate. There have been many attempts to overcome the asymmetry problem such as the LSB matching (LSBM) [7] but with limited success, distortions may occur undermining the secrecy of hidden data or help estimate its size [8].

Attempts to design steganography schemes that generate less distortion has motivated the introduction of the concept of embedding efficiency in [9] which was first adopted in [5] for embedding in DCT domain. For the LSBR or LSBM schemes, the probability of pixel change is 0.5, i.e. embedding efficiency of 2, [5]. The embedding efficiency attribute directly influence security, because smaller number of embedding changes is less likely to disrupt statistic properties of the cover image [10]. Ker *et al.* listed the control of embedding efficiency as one of the future research challenges in steganography [11].

So far, research into designing steganography schemes with high embedding efficiency rather limited. The *matrix encoding* technique proposed by [9] was the first attempt to improve LSB based embedding efficiency. It embeds $k$ secret bits by manipulating a group of $2^k - 1$ cover image pixels' LSB but changing at most one pixel. On average, when $k = 2$, 25% of pixels are changed but payload capacity is limited to 67%. In general, embedding $k$ bits using this method, increases embedding efficiency to $2^k$ but limits the capacity to $k/(2^k - 1)$. Thus, such kinds of embedding techniques are not useful for those applications that require high payload capacity.

Several schemes have been designed with the objective of increasing the embedding efficiency that adopt the same strategy of the matrix encoding to embed several secret bits in a number of contiguous pixels. The most common feature in these schemes is the use of one or more binary functions defined on a set of 2 or 3 neighboring cover pixels designed to avoid changing more than one of the neighboring pixels are changed. These schemes include: 1) the LSB matching revisited (LSBMR) proposed as a variant of LSBM by Mielikainen, [12]; 2) Chan's modification of the LSBMR, [13]; and 3) the Iranpour *et al.* scheme, [14], which generalizes both previous schemes and embeds 3 secret bits in 3 consecutive pixels using 3 similarly defined binary functions but in rare cases all 3 cover pixels could change. Theoretically, for the first and last schemes, the binary function(s) adopted in these schemes reduce the probability of change from 0.5 to 0.375 with EE=2.66, but in many cases at the expense of lower capacity. The lower capacity is due to fact that saturated pixel values {0, 255} are not used for secret embedding.

In [15], the authors explored the idea of increasing similarity between secret image bit-stream and cover image LSB plane for improved efficiency. A compression-like algorithm, called *secret image size reduction* (SISR), was shown to reduce the secret image bit-stream length by approximately 30% without losing information, and whereby 57% of the bits in the output bit-stream have a 0 value. On the other hand, decomposing cover image pixel values using Fibonacci scheme, instead of traditional binary, produced 61% of 0's in the cover image LSB. Consequently, embedding SISR secret bit-stream in the Fibonacci decomposed cover image LSB plane results in increased embedding efficiency compared to LSBR scheme. The rest of the paper is devoted to extend and further refine the idea of increasing similarity between the secret image bit-stream and the cover image LSB plane to further increase embedding efficiency while maintaining capacity. We shall also test robustness of the proposed steganography schemes against 3 well-known steganalysis tools: the *revised weighted stego* (RWS) [16] which is an improvement version of the *weighted stego* (WS) [17]; the *difference image histogram* (DIH) [18]; and the *LSB matching steganalyser* (LSBMS) [19]).

## III. PROPOSED METHODS

This section is devoted to develop image processing procedures and models that can be used to achieve a high similarity between secret image bits and the cover pixels' LSB by increasing the ratio of 0's for both secret image bit-stream and cover image LSB plane. We propose a spatial domain as well as a wavelet domain algorithm to transform secret bit-streams to significantly increase ratio of 0's. We shall also investigate cover pixel value decomposition schemes that increase the ratio of 0's in the cover image LSB plane.

### A. Secret Image Manipulation (SIM)

This algorithm uses the secret image histogram to define a grayscale transform that maps secret image pixel values according to the descending order of their frequencies so that more frequent pixel values are mapped into bytes with lower number of 1's. For that we first partition the set of all possible 256 grayscale values into 9 ordered subset $S_i$, i=0,…,8, consisting of all 8-bit strings with exactly i of 1's in ascending order of their decimal values, see Table I, below. When two or more pixel values have the same frequencies, then map according to appearance in the sorted frequency vector. This transform is simply a substitution function on the histogram of the secret image, with no loss of information.

*1)* The SIM Transforming Procedure

1- Obtain the histogram $h$ of the secret image $I$.
2- Let $h'$ be $h$ in descending order of frequency.
3- Based on $h'$, replace the most frequent pixel value in $I$ with the first new value in the Table I, and continue by replacing the next highest repeated pixel value by the next new value. This results in a new image $I'$.
4- Covert $I'$ into the bit-stream.
5- Construct a side information $(9+8\,N)$ bit-stream, where $N$ refers to the number of pixel values present in $I$. The first 9 bits of the side information represent $N$, and the next $8\,N$ bits list the original pixel values in descending order of frequencies.
6- Append the bit-stream of the secret image $I'$ to the side information bit-stream, and embed in the chosen cover image using the given hiding scheme.

Note that the maximum possible number of bits for the side information part is (9 + 256 * 8 = 2057). This will reduce the payload capacity, but only by a very negligible proportion. The second part of the bit-stream represents the modified secret image $I'$. Fig. 1 below displays a secret image $I$ (Lenna) and its SIM modified version $I'$.

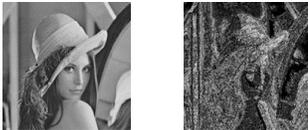

Fig. 1. Lenna image and its modified version using SIM algorithm.

*2)* The SIM Reconstruction Procedure

1- Extract the side information and the SIM modified secret image $I'$, and let $h'$ is the histogram of $I'$.
2- The original image $I$ is recovered by replacing the pixel values in the image $I'$ that has the i[th] value in the $h'$ with the i[th] value of the reconstructed original pixel values from the side information.

*3)* Performance of SIM

To test the performance of the SIM algorithm in terms of percentage of 0's in secret image bit-stream before and after SIM substitution, we use images from the following databases:
1. USC-SIPI database, [20]. We use 44 grayscale images from the Miscellaneous volume which consists of 16 colour images and 28 monochrome images. These include some known images such as Lenna, Baboon, Peppers, Jet, Tiffany, Couple, Bridge, Pirate, House and Lake. We use 3 different sizes (512 x 512, 256 x 256, and 128 x 256).
2. BOSSBase version 1.0 database of 512 x 512 grayscale images, [21]. We select the first 1000 images out of the 10000 images that include, but not limited to, landscapes, people, plants, and building. We resized images to 256 x 256, and 128 x 256.

In Table II and Table III, below, we present the results of the experiments for the SIPI and BOSSBase databases, respectively. Each table presents 4 parameters (mean μ, standard deviation σ, minimum $M_n$, and maximum $M_x$) of the percentage of 0's in the secret image bit-streams before and after SIM substitution as well as the corresponding length of the side information. Here $R$ refers to the original images while $R'$ refers to the post SIM images, and $L$ refers to the length of the side information. From Table II, we note that on average the percentage of 0's in the SIM modified images is increased by about 45% over that in the original images. Similarly, the results of Table III show an increase of 26% in percentage of 0's post SIM. The difference between the rates of increase reflects the variation in the nature of images in the two databases. The statistical parameters (μ and σ) in Table III are independent of image size, but in Table II different size images result in marginal variation in these parameters. This variation cannot be attributed to the effect of the SIM.

The only drawback of the SIM is the need for embedding the side information which results in slight decrease in embedding capacity. However, the results in tables II and III reveal that the average proportion of the side information to the SIM-modified secret image bit-streams is negligible and diminishes for larger size images (size 128x256: 0.006 (SIPI), and 0.007 (BOSSBase); size 256x256: 0.003 (SIPI), and 0.004 (BOSSBase); size 512x512: 0.001 for both dataset).

The SIM algorithm doesn't yield similar performance if applied to non-image secret bit-streams. This is due to the fact that, unlike secret image bit-streams, the frequency distributions of 8-bits bytes extracted from non-image secret bit-streams are highly likely (or for security reasons are expected) to be uniform. In the next section we will develop an Integer Wavelet domain version of SIM with higher performance in terms of percentage of 0's.

TABLE I
PARTITIONING GRAYSCALE VALUES IN TERMS OF NO. OF 1'S.

| S | $S_0$ | $S_1$ | $S_2$ | $S_3$ | $S_4$ | $S_t$, t=5,…,8 |
|---|---|---|---|---|---|---|
| PIXEL VALUES | {0} | { $2^I$: I=0,…,7} | { $2^I+2^J$: I=0,…,6 AND J=I+1,…,7} | { $2^I+2^J+2^K$: I= 0,…,5, J=I+1,…,6, K=J+1,…,7} | { $2^I+2^J+2^K+2^L$: I= 0,…,4, J=I+1,…,5, K=J+1,…,6, L=K+1,…,7} | {255 - s:  s ε $S_{8-t}$ } |

TABLE II
SIPI DATABASE – EFFECT OF SIM ON THE PERCENTAGE OF 0'S

| | Image size 128 x 256 | | | Image size 256 x 256 | | | Image size 512 x 512 | | |
|---|---|---|---|---|---|---|---|---|---|
| | $R$ | $R'$ | $L$ | $R$ | $R'$ | $L$ | $R$ | $R'$ | $L$ |
| µ | 0.49 | 0.71 | 1639 | 0.49 | 0.71 | 1676 | 0.49 | 0.73 | 1565 |
| σ | 0.08 | 0.07 | 279 | 0.09 | 0.07 | 273 | 0.10 | 0.08 | 499 |
| $M_n$ | 0.12 | 0.61 | 993 | 0.12 | 0.60 | 993 | 0.11 | 0.60 | 25 |
| $M_x$ | 0.65 | 0.93 | 2057 | 0.66 | 0.94 | 2057 | 0.66 | 0.99 | 2057 |

TABLE III
BOSSBASE DATABASE - EFFECT OF SIM ON THE PERCENTAGE OF 0'S.

| | Image size 128 x 256 | | | Image size 256 x 256 | | | Image size 512 x 512 | | |
|---|---|---|---|---|---|---|---|---|---|
| | $R$ | $R'$ | $L$ | $R$ | $R'$ | $L$ | $R$ | $R'$ | $L$ |
| µ | 0.54 | 0.68 | 1815 | 0.54 | 0.68 | 1850 | 0.54 | 0.68 | 1912 |
| σ | 0.07 | 0.05 | 278 | 0.07 | 0.05 | 260 | 0.07 | 0.05 | 228 |
| $M_n$ | 0.17 | 0.57 | 385 | 0.16 | 0.57 | 537 | 0.15 | 0.56 | 777 |
| $M_x$ | 0.85 | 0.93 | 2057 | 0.86 | 0.93 | 2057 | 0.86 | 0.92 | 2057 |

*B. Integer Wavelet based SIM (IWSIM)*

Discrete wavelet transforms (DWT) provide a multi-resolution representation of signals at a different scale as well as different frequency ranges (sub-bands). There are a variety of wavelet filter banks to use for decomposing an image into different frequency sub-bands. At scale level 1, a wavelet decomposition scheme partition any image into four sub-bands, namely Low-Low (LL) sub-band, Low-High (LH), High-Low (HL) and High-High (HH) sub-bands. At subsequent levels, the current LL sub-band is decomposed again into 4 sub-bands. In all sub-bands, the wavelet coefficients are real numbers, but integer wavelet transforms (IWT) can be defined in several ways [22]. Here we use the integer version of Haar wavelet.

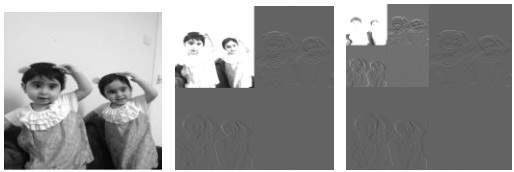

Fig. 2. An image and its 2-Level Wavelet Decomposition

The main incentive for extending the SIM into the DWT domain is the fact that the histogram of the LL sub-band is an approximation of that of the original image, while the coefficients in the high frequency sub-bands have a Laplacian distribution. However, the traditional IWT decomposed image may contain coefficients outside the usual grayscale range of [0...255]. At level 1 decomposition, the high frequency LH, HL, and HH sub-bands coefficients may require up to 10 bits to represent. Therefore, we need to modify the SIM mapping and adjust the side-information accordingly. At IWT decomposition level 2 or above, coefficients ranges usually expand and require even more than 10-bits to represent. This is why we apply the IWT only to level one, because at higher level decomposition requires much larger SIM-like mapping tables and increased size of the side-information that would reduce embedding capacity. Another reason for avoiding level 2 of IWT is the number of sub-bands increase to 7 sub-bands, and then each sub-band needs its own side-information, and this again reflects on increasing the total side-information size for the secret image.

In the same way that Table I was constructed, the modified mapping depends on partitioning of the set of 9-bits strings into 10 subsets $S_i$, consisting of all bit-strings that exactly have i of 1's arranged in ascending order of their decimal values. The modified mapping of the wavelet coefficients onto the bit-strings in these partitions follow the same SIM theme of mapping most frequent coefficients to the remaining bit-strings with the lowest number of 1's.

For *LL* sub-band, IWSIM simply applies the SIM procedure, but for the other sub-bands, IWSIM uses the modified mapping and constructs 2 different format side-information (IWSIM1 and IWSIM2) depending on whether the number of distinct coefficients in the sub-band is ≤ 256 or not. To avoid dealing with negative coefficients we transform the sub-bands by subtracting the minimum coefficient value.

*1) The IWSIM Transforming Procedure*

Decompose the secret image using Haar IWT. For the *LL* sub-band call the SIM procedure. For each other sub-band $S$ (LH, HL, and HH) follow the steps below:

1- Calculate $m = \min(S)$.
2- Let $S' = \{s - m : s \, \varepsilon \, S\}$.
3- Compute the histogram $h$ of $S'$.
4- Let $h'$ be $h$ in descending order of frequency.
5- Let $C$ be the number of different coefficients in $S'$.
6- If ($C \leq 256$), construct the IWSIM1 side-information. Else construct the IWSIM2 side-information.
7- Use $h'$ to map $S'$ onto the corresponding modified SIM-like partitioned bit-strings. This yields a new sub-band $S''$.
8- Form the secret bit-stream by concatenating the bit-strings in $S''$.
9- Append secret bit-stream to the side-information as defined below.

The *IWSIM1 side-information $SI_1$* is the concatenation of 8 –bits for m, 2 bits as indicator of the length of the bit-string needed to represent coefficients in $S'$ (00, 01, or 10 for 8, 9, or 10 bits resp.), 9 bits to represent $C$, and $C$x(8 or 9 or 10) bits as required to list the $S'$ values in descending order of frequencies.

The *IWSIM2 side-information $SI_2$* is the concatenation of 8 –bits for m, 1 bit as indicator of the length of the bit-string needed to represent coefficients in $S'$ (0 for 9, or 1 for 10), 10 bits to represent $C$, and $C$x(9 or 10) bits as required to list the $S'$ values in descending order of frequencies.

Finally, append $SI_i$, i=1,2, to a 1 bit indicator set to i-1. Note that, max{length ($SI_1$) = 1+(8+2+ 9+(256x10) = 2580) bits, while max{length ($SI_2$) = 1+(8+1+10+(512x10) = 5140) bits.

*2) IWSIM Reconstruction Procedure*

After extracting the secret bit-stream, in accordance with the given embedding scheme, the first bit indicates whether one reads the remaining 19 as $SI_1$ (i.e. 19=8+2+9) or as $SI_2$ (i.e. 19=8+1+10). The various parts of the 19-bits determine the value of m, the coefficient binary string length, and the value of $C$. The rest of the extracted bit-stream is $S''$. The original sub-band $S$ can be reconstructed from $S''$ as follows:

> 1- Determine the histogram $h'$ of $S''$.
> 2- Construct $S'$ by replacing the $S''$ sub-strings that has the i$^{th}$ value in $h'$ with the i$^{th}$ value of the reconstructed values from the side-information.
> 3- Let $S = \{s + m: s \; \varepsilon \; S'\}$.

Finally, after extracting *LL*, *HL*, *LH*, and *HH* subbands use Inverse IWT to reconstruct the original secret image *I*.

*3) Performance of IWSIM*

To test the performance of the IWSIM algorithm in terms of percentage of 0 bits in the secret image bit-stream before and after modification, we repeated the experiments, when tested the SIM with 2 databases and image sizes. Results of the experiments conducted for three different image sizes are shown in Table IV and Table V for the SIPI and BOSSBase databases, respectively. We note that on average, percentage of 0's in the IWSIM bit-streams is increased by about 66% (for SIPI) and 49% (for BOSSBase) over that of the original images. As in the case of the SIM procedure, these results demonstrate that the performance of the IWSIM is more influenced by the nature of the secret image than by the image size. Furthermore, these results also demonstrate that wavelet decomposition of secret images in the IWSIM leads to significant increase in the percentage of 0's in the secret bit-stream for both databases over the SIM algorithm. The improvement is less significant for the SIPI database (around 47%) than for the BOSSBase database (around 89%). This improvement is a consequence of the fact, mentioned earlier, that the wavelet coefficients in the 3 high frequency sub-bands have Laplacian (generalized Gaussian) distributions with 0 means and low standard deviations. On the other hand, distribution of the secret image spatial domain pixels could vary from one image to another. The nearer the secret image histogram is to a Gaussian distribution, the nearer the performance of the SIM is to that of the IWSIM.

As in the case of the SIM algorithm, there is a need to add the side-information. Although the length of the IWSIM side-information is on average about 4 times the SIM side-information, it is equally negligible in comparison to the size of the secret bit-stream. In fact, the increase in the length of side-information as a result of using SIM (resp. IWSIM) only limits the embedding capacity to 99.3 % (resp. 98%) compare to the capacity of the usual LSB embedding scheme.

TABLE IV
SIPI DATABASE – EFFECT OF IWSIM ON THE PERCENTAGE OF 0'S

|  | Image size 128 x 256 | | | Image size 256 x 256 | | | Image size 512 x 512 | | |
| --- | --- | --- | --- | --- | --- | --- | --- | --- | --- |
|  | R | R' | L | R | R' | L | R | R' | L |
| µ | 0.49 | 0.81 | 5171 | 0.49 | 0.81 | 5358 | 0.49 | 0.84 | 5238 |
| σ | 0.08 | 0.06 | 1590 | 0.09 | 0.06 | 1607 | 0.10 | 0.06 | 2272 |
| $M_n$ | 0.12 | 0.73 | 1188 | 0.12 | 0.73 | 1268 | 0.11 | 0.76 | 334 |
| $M_x$ | 0.65 | 0.99 | 11054 | 0.66 | 1.00 | 10629 | 0.66 | 0.99 | 11212 |

TABLE V
BOSSBASE DATABASE - EFFECT OF IWSIM ON THE PERCENTAGE OF 0'S.

|  | Image size 128 x 256 | | | Image size 256 x 256 | | | Image size 512 x 512 | | |
| --- | --- | --- | --- | --- | --- | --- | --- | --- | --- |
|  | R | R' | L | R | R' | L | R | R' | L |
| µ | 0.54 | 0.80 | 5277 | 0.54 | 0.81 | 5829 | 0.54 | 0.83 | 6781 |
| σ | 0.07 | 0.03 | 1286 | 0.07 | 0.03 | 1432 | 0.07 | 0.04 | 1781 |
| $M_n$ | 0.17 | 0.70 | 1220 | 0.16 | 0.72 | 1644 | 0.15 | 0.74 | 1756 |
| $M_x$ | 0.85 | 0.94 | 9081 | 0.86 | 0.95 | 10039 | 0.86 | 0.97 | 12015 |

*C. Pixel value decomposition schemes*

In this section we focus on investigating different representations of (cover) images that can produce higher proportion of 0's than 1's in their LSB bit-plane. The objective is to increase the probability of similarity between the bits value in the secret image bit-stream (post SIM or IWSIM) and the cover pixels' LSB plane, and consequently significantly improve the embedding efficiency of the usual LSBR.

In recent years, image pixel-value decomposition schemes other than the conventional binary scheme have been proposed for use in steganography primarily to increase payload capacity by embedding in bit-planes beyond the LSB without adverse impact on stego-image quality. These schemes include the Fibonacci [23], prime [24], natural [25], Lucas [26], Catalan-Fibonacci (CF) [27]and the Simple Sequence (SS) [28]. Interestingly, these schemes can be shown to increase the ratio of 0's in LSB plane but at the expense of reduced embedding capacity. In the rest of this section we describe the way such schemes work highlighting the pros and cons, introduce a new pixel value decomposition scheme (which we call the *Extended-Binary)* that avoids some of the shortcomings, and demonstrate that it outperforms all the above schemes, except the so called natural scheme, in terms of the ratio of 0's in the LSB plane.

*1)* Background

The intensity values of grayscale images range from 0 to 255 requiring 8 bits to be uniquely represented in binary in terms of the defining sequence $\{1, 2, 2^2, 2^3, …,2^7\}$. The Fibonacci, prime, Lucas, CF, natural and SS pixel value decomposition schemes offer other longer defining sequences to represent grayscale values in 12, 15, 12, 15, 23, and 16 bits respectively. However, unlike the binary decomposition scheme, non-binary decomposition schemes do not result in a unique bit-stream representation of pixel values. For example, the Fibonacci scheme encodes the grayscale value 5 as 000000001000 or 000000000110. Uniqueness is enforced by not allowing the use of consecutive Fibonacci numbers. In this case 000000000110 is not a valid Fibonacci code for 5 based on Zeckendorf's theorem.

<u>*Zeckendorf's Theorem*</u>: *Each positive integer can be represented as the sum of distinct, but not consecutive, Fibonacci numbers*.

For the other non-binary decomposition schemes uniqueness is imposed by selecting bit-string codes of lexicographically highest value.

The inclusion of some odd numbers in the defining sequences of these schemes together with the restrictions that need to be imposed to guarantee unique decomposition (e.g. the Zeckendorf theory) will be shown later to increase the ratio of 0's in the resulting LSB plane. However, this is achieved at a price that limits their benefits for steganography. While the uniqueness is solved by the above theorem, Fibonacci based embedding technique faces another problem in that the very act of embedding could result in violating the Zeckendrof condition. For example, using LSBR to embed a secret bit 1 in the valid Fibonacci code of the cover pixel value 7 = (000000001010) yields stego pixel value (000000001011) which cannot be recognized at the receiving end as valid Fibonacci code and even if the embedding scheme is modified to replace it with the Fibonacci code (000000010000) the receiver wrongly extracts a 0 secret bit. To avoid such a situation, the Fibonacci steganography scheme skip cover pixels for which embedding certain secret bits cause a violation of the Zeckendorf condition. To retrieve the secret data, the selected stego pixel value is first decomposed into Fibonacci representation, and then it needs to be checked whether it is a good candidate or not, if it is, then the secret bit is extracted from the agreed bit-plane. As a result the embedding capacity is degraded (note that the same problem is facing other non-binary schemes except SS). To overcome the capacity limitation, embedding in other than the LSB plane has been proposed, without avoiding payload capacity limitation. In [29] and [15], this problem was considered and an innovative solution was proposed by using mapping table. The idea of mapping based embedding technique was first suggested by [29]to embed two secret bits in each Fibonacci decomposed cover pixel value, and then extended in [15] to embed one secret bit in each Fibonacci decomposed cover pixel value to improve the stego-image quality.

*2)* The Extended-Binary decomposition scheme

The defining sequence *K* of any pixel value decomposition scheme includes {1}. Increasing 0's ratio for any decomposition scheme is only possible if odd pixel values can be expressed without using 1. This is not the case for the usual binary decomposition scheme, or any scheme whose defining sequence does not include any odd number >1. Expanding the usual binary scheme (B) by adding an odd integer $x$ >1 can help but the level of increase depends on the distribution of image pixel values as well as the value of *x*. Given an image *I* of size *N,* let hist(*I*) be its histogram. The amount of increase of ratio of 0's in the LSB plane as a result of adding an odd number *x* to B is dependent on hist(*x*) and hist(*y*) for all $y > x$ that can be expressed by the extended scheme without using 1. Consequently, *x* should be relatively small. Next we discuss our approach to selecting *x*.

Let B = $\{1, 2, 2^2, …,2^7\}$of the usual binary scheme, and *x* be an odd number < 256. Note that, uniqueness of representation with respect to $B \cup \{x\}$, requires the use of lexicographically highest decomposition. For any *x*, many odd numbers > *x* will have 0 LSB. However, several odd numbers do not have their LSB changed and worse some even numbers will have 1 LSB. For example, if *x* =5 then the even numbers in the set $\{6 + 8i\}_0^{31}$ as well as the odd numbers in the set $\{1 + 8i\}_0^{31} \cup \{3 + 8i\}_0^{31}$ all have 1 LSB. On the other hand, if *x* is of the form $2^i$-1, then no even number changes its LSB.

The smallest odd number of the form $2^i$ - 1 is *x* =3 and in this case all odd numbers of the form$\{3 + 4i\}_0^{63}$, will have 0 LSB, whereas the LSB of all other odd numbers remain as 1.

In the rest of the paper we adopt the *Extended-Binary* pixel decomposition scheme *S* with the defining sequence:

$$S = \{3\} \cup \{2^n \mid 0 \leq n \leq 7\} \quad (2)$$

*3)* Performance of Extended-Binary scheme (S)

Only 50% of the all possible 256 grayscale values have 0 LSB when decomposed by the binary scheme. The discussion above show that 192 out of the 256 grayscale values, have 0 LSB when *S* is used, i.e. 75% of the grayscale values have 0 LSB. However, the effect of any decomposition scheme on the LSB bit-plane of a specific image *I* depends on the distribution of its pixel values. The extended binary scheme *S* increases the ratio 0's in the LSB of *I* by:

$$\frac{\sum_{i=0}^{63} hist\ (3 + 4 * i)}{N} \qquad (3)$$

In this section, we shall demonstrate that *S* outperforms almost all known pixel decomposition schemes in terms of ratio of 0 LSB when applied to real images. We shall use the 512 x 512 size images from the SIPI and BOSSBase databases that we used for testing SIM and IWSIM. We first, compare the performance of *S* with several other extended binary schemes whose defining sequences differ from that of *S* only in the added odd number *x*. In particular, we consider 5 extended sequences $S_1, S_2, S_3, S_4,$ and $S_5$ by adding the 5 prime numbers 5, 11, 23, 47 and 97, respectively.

Fig. 3 presents the average ratio of 0 LSB achieved by the six Extended-Binary schemes described above. The results confirm the superiority of the *S* scheme in terms of the 0 ratios and also show that even the worst Extended-Binary scheme performs better than the usual binary scheme which only achieves around 47% ratio of 0 LSB for the tested images.

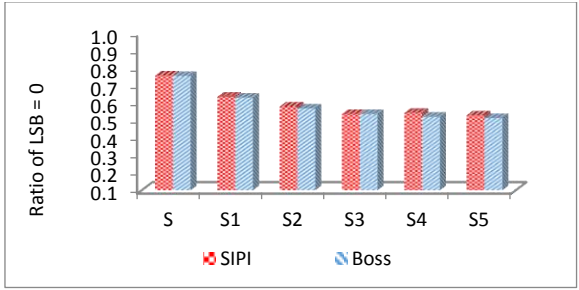

Fig. 3. Ratio of 0 LSB of cover pixels for different Extended Binary schemes.

For any decomposition scheme, one can split the grayscale interval into 4 subsets $A_{ij}$ for i,j ε {0,1}, representing the values whose LSB is i and the scheme makes it j. For any specific image *I*, the overall effect of any decomposition scheme on the ratio of 0 LSB depends on the frequency distribution of these 4 subsets of the grayscale interval. The increase in ratio of 0 LSB for image *I* is depedent on the value of ($hist(A_{10}) - hist(A_{01})$). For our scheme *S*,

$$A_{00} = E,\ A_{01} = \Phi,\ A_{10} = \{3+4k: k=1,\ldots, 63\},\ \text{and}$$
$$A_{11} = O - A_{10}.$$

Here, *E* (respectively *O*) is the set of all even (respectively odd) grayscale values and $\Phi$ is the empty set. Note that for all the other Extended-Binary schemes above, $A_{01} \neq \Phi$, and $A_{10}$ has much fewer values than 64. In fact, only when one extends the binary scheme with an odd number of the form $2^i -1$ one has $A_{01} = \Phi$. These facts seem to give the *S* scheme more chances to outperform the other schemes but only if we assume reasonable distribution of the coefficients in $A_{10}$ and $A_{11}$. In fact in all the experiments we note that there were noticeable variations away from the averages shown in Fig. 3.

Fig. 4 presents the results of the same experiments but to compare the performance of our scheme *S* with the other decomposition schemes mentioned earlier. Again the testing was carried out on the selected 512x512 images from SIPI and BOSSBase databases.

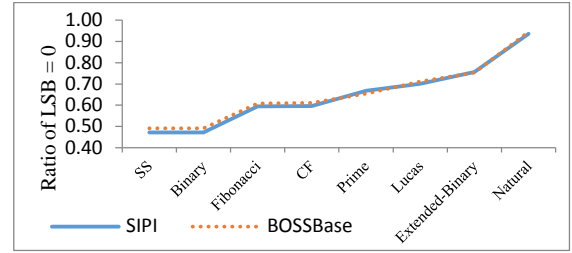

Fig. 4. Ratio of 0 LSB for different decomposition techniques.

While the highest ratio is obtained when using the natural scheme followed by our *S* scheme. The *SS* scheme does not increase the ratio of 0 LSB over the binary scheme due to absence of odd numbers >1 resulting in $A_{10} = \Phi$. The performance of the other schemes can be explained by examining their 4 $A_{ij}$ sets and their histograms in cover images. In all cases one can easily find that $A_{01} \neq \Phi$. For the Fibonacci scheme, the ratio of 0 LSB is significantly influenced by the frequency of the 51 values in $A_{01}$ which reduces the influence of the 80 values in $A_{10}$.

Further increase in the ratio of 0 LSB may be possible for some images if invertable image transformation can be found that boost the histogram of $A_{10}$ and/or degrade the histogram of $A_{10}$. In the next section, we consider the use of the image complement transform.

### D. The Proposed Mapping-based embedding technique

The work in the last two sections results in achieving high similarity between secret image bit-streams and the LSB plane of cover images which can reduce the amount pixel changes for LSB based steganography. However, the success in increasing the ratio of 0 LSB using a non-binary decomposition schemes could have a drawback by rendering some pixel unsuitable for hiding a secret bits and thereby reducing payload capacity. We shall now, describe our modification of the LSBR that uses binary mapping table for embedding secret bit-streams that aceives nearly full capacity for certain combinations of cover pixel decomposition and secret image bit-stream manipulation schemes. This strategy builds on our earlier work with the modified Fibonacci scheme, [29] and [15].

*1) Embedding Mapping Table for Non-binary schemes*

Besides increasing the ratio of 0 LSB bits in cover image pixels, the non-binary decomposition schemes reduce the number of possible patterns in the lowest 3 bit-planes in the decomposed cover pixels. For these decomposition schemes the number of possible 3-bit patterns is reduced into 4 or 5 out of 8 different random patterns. However, in all but 3 of our investigated schemes changing the LSB as a result of embedding a secret bit may result in breaking the uniqueness property. The 3 schemes that allow using embedding mappings are the Fibonacci, the Lucas and our Extended-Binary scheme *S*.

TABLE VI
MAPPING FOR FIBONACCI, LUCAS, AND S

| COVER 3-LSB | FIBONACCI SECRET BIT | | LUCAS SECRET BIT | | S SECRET BIT | |
|---|---|---|---|---|---|---|
| | 0 | 1 | 0 | 1 | 0 | 1 |
| 000 | 000 | 001 | 000 | 001 | 000 | 001 |
| 001 | 010 | 001 | 010 | 001 | 010 | 001 |
| 010 | 010 | 001 | 010 | 001 | 010 | 001 |
| 100 | 100 | 101 | 100 | 101 | 100 | 001 |
| 101 | 100 | 101 | N/A | | | |

The above table shows that, for the 3 schemes, all pixel values become feasible for full capacity embedding and the embedded secret bits are simply the LSB in the decomposed cover image. At the receiver part, the secret bits can be extracted from the LSB of the stego-image pixels. For the Fibonacci and the S schemes embedding, a single bit may increase/decrease cover pixel value by a maximum of 1, while for the Lucas scheme the maximum change is 2, because the first element in the Lucas sequence starts by 2. The impact of these potential changes on the stego-image quality depends on the distribution of the various 3-bit patterns in the decomposed cover image pixels. In what follow we use these single-bit mapping for LSB-like steganography although it can be modified for other schemes.

*E. Embedding and extracting procedure*

By combining each of our two secret image pre-processing algorithms with the pixel Extended-Binary decomposition scheme and using the corresponding mapping table, we get two different schemes that referred to by EB_SIM, and EB_IWSIM. Here, we shall present a general format of the embedding and extracting procedures for each possible paired scheme.

*1) Embedding Procedure*
  1. Apply the SIM, or IWSIM on the secret image prior to embedding producing the secret bit-stream of length $m$.
  2. Let $I'$ be the complement image of the cover image $I$.
  3. Decompose pixels value using the S-version of the Extended-Binary decomposition technique for $I$ and $I'$.
  4. Calculate the 0 ratio $R$ and $R'$ of the LSB plane of the decomposed image $I$ and $I'$, respectively.
  5. If $R >= R'$, then the image $I$ is chosen as a cover, otherwise, image $I'$ is chosen as a cover.
  6. PRNG is used to select the cover pixel $p_i$ randomly to be used for message embedding using an agreed seed.
  7. Based on the proposed mapping in Table VI, the secret bit $m_i$ is embedding in $p_i$.

Note that one bit is needed to be added to the secret bit-stream to indicate to the receiver whether the secret is embedded in the decomposed version of $I$ or in that of $I'$. In the first case, the bit is set to 0 otherwise it is set to 1.

*2) Extracting Procedure*
On receiving the perceived stego-image $S$, first the indicator bit should be extracted from the agreed pixel location.
  1. If the indicator bit is 0, then extract the secret from $S$, else extract it from the complement image $S'$.
  2. Use same PRNG to select the random stego pixel $p_i'$.
  3. Extract the secret bit $m_i$ from the LSB of the Extended-Binary representation of $p_i'$ using the appropriate mapping table.
  4. The reverse procedure (decoding) of the SIM, or IWSIM is applied on the extracted bit-stream to reconstruct the embedded secret image.

For comparison reasons, we also create two other mapping based embedding schemes using the above procedures for the IWSIM pre-processing but instead of the Extended-Binary the cover images will be decomposed by Fibonacci and Lucas. We refer to these schemes as Fib_IWSIM and L_IWSIM.

IV. EXPERIMENTAL SETUP AND RESULTS

In this section, the performance of the proposed image-based steganography schemes (EB_SIM, EB_IWSIM, Fib_IWSIM, and L_IWSIM) is evaluated and compared with the performance of the LSBR, LSBM, and LSBMR steganography schemes. A more extensive evaluation was conducted and presented in [30]. These evaluations aim to: 1) measure payload capacity, 2) measure embedding efficiency, 3) test stego-image quality, and 4) measure the detectability/security of the embedded message. The results are in all these tests consists of the 44 images from SIPI and the first 1000 images from BOSSBase as cover 512x512 images. For secret images, we embedded the 44 SIPI images but resized to 128 x 256 to be embedded in each of the 44 SIPI cover images, resulting in 1936 stego-images. In each BOSSBase cover images we embed the Lenna image of size 128 x 256 as a secret resulting in 1000 stego-images. The evaluation will be conducted for secret bit-streams embedding at 5 different percentage of the available capacity, namely 20%, 40%, 60%, 80% and 100%.

*A. Payload Capacity Evaluation*

The payload capacity is measured as the proportion of the embedded secret bits to the cover image size. Table VII, shows the results of test (1) for all tested steganography schemes.

TABLE VII
CAPACITY OF THE TESTED STEGANOGRAPHY SCHEMES.

| | SIPI | BOSSBase |
|---|---|---|
| LSBR | 1.0 | 1.0 |
| LSBM | 1.0 | 1.0 |
| LSBMR | 0.952 | 0.978 |
| EB_SIM | 0.994 | 0.993 |
| EB_IWSIM | 0.978 | 0.979 |
| Fib_IWSIM | 0.978 | 0.979 |
| L_IWSIM | 0.978 | 0.979 |

While the LSBR and LSBM achieve full capacity, our EB_SIM is only marginally lower. The lowest average capacity (0.952) is achieved by the LSBMR for the SIPI database. In all other cases, a capacity of around 0.98 is achieved. The loss in capacity by the LSBMR technique is entirely due to the exclusion of the saturated cover pixel values (i.e. 0 or 255) which account for an average of 4.8% for the SIPI images and 2.2% for the BOSSBase database. Whereas the loss capacity in the cases of EB_SIM, EB_IWSIM, Fib_IWSIM, and L_IWSIM is accounted for by the size of the side-information appended to the actual secret, and in the case of IWSIM-based there is an increase in the number of bits representing coefficients in some Wavelet sub-bands.

### B. Embedding Efficiency Evaluation

Theoretically, the probability ratio of cover pixels that would change post embedding is proportional to the embedded secret image size, and is calculated as:

$$p = 1 - (R_0 \times R'_0) + ((1- R_0) \times (1- R'_0)) \qquad (4)$$

where $R_0$ is the ratio of 0's in the secret bit-stream, and $R'_0$ is the ratio of 0's in the LSB of the cover pixels. At full embedding capacity, on average IWSIM achieves 80% ratio of 0's in the secret bit-streams while the LSB plane of the $S$ decomposed cover images contain 77% of 0's. Therefore, excluding the effect of the side-information, the EB_IWSIM is expected to acheive an embedding efficiency $p = 0.338$ (EE=2.96) compared to the LSBR scheme for which $p = 0.5$ (EE=2).

In practice, the value of EE of our proposed embedding schemes depends on: 1) the amount of similarity between the ratio of 0's in the secret bit-stream and the ratio of 0 LSB of the decomposed cover image, 2) the size of the side-information. Figures 5 and 6, below, present the charts showing average EE values at the different embedding proportions obtained from all the stego-images obtained from the SIPI and BOSSBase databases, respectively.

From these charts, one can see that the EB_IWSIM outperforms all other schemes for the payload of 60% or more, but it is outperformed by the LSBMR at the lower embedding rates. For the EB_IWSIM, and all embedding schemes that use IWSIM secret image transformation, the EE value decreases as embedding rate decrease which can be attributed to the effect of including the side information. Note that, the size of the side-information is a constant across different embedding rates. The performance of the EB_SIM is the lowest among all our proposed schemes but is independent of the embedding rate. The rather impressive performance of LSBMR, with almost fixed EE at all rates, provide a strong incentive to develop and investigate new versions LSBMR that could apply post pre-processing of secret bit-streams and cover pixel decompositions.

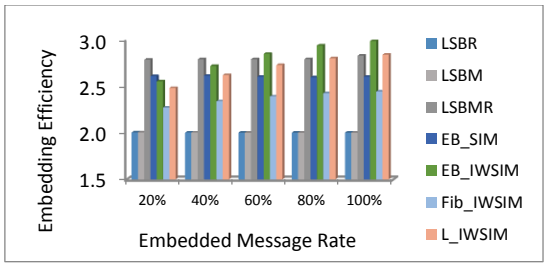

Fig. 5. Embedding efficiency for the SIPI database.

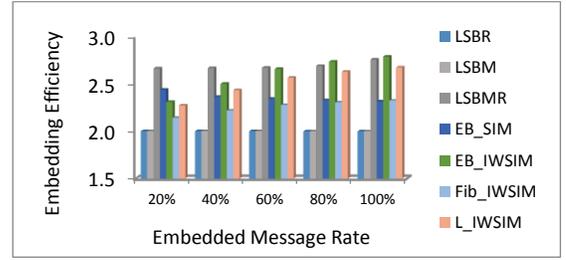

Fig. 6. Embedding efficiency for the BOSSBase database.

### C. Stego-Image Quality Evaluation

We evaluated the stego-image quality for all the above 7 embedding schemes in terms of the average PSNR values with respect to the original cover images. The results for the SIPI database are shown in Fig. 7. The PSNR of the stego-images in the BOSSBase, have identical patterns with marginally lower averages and hence are not shown.

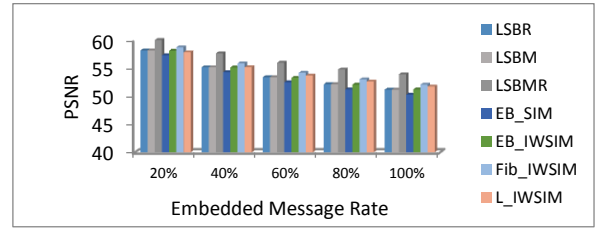

Fig. 7. Average PSNR for the tested schemes - SIPI database.

These charts confirm that stego-images generated by our schemes are of reasonable quality comparable to those output by LSBR and LSBM. The stego-images output by the LSBMR scheme are of higher quality. The Fib_IWSIM performance is reasonably near that of the LSBMR. Note that 25% of the lowest 3 bit-planes of the EB decomposed cover pixels are 100 and if the secret bit value is 1 then the cover pixels value will change by 2.

### D. Detectability Evaluation

In this section, we report on experiments conducted to test the robustness of our mapping-based embedding schemes against 3 steganalysis detectors that are commonly used to detect the presence of secrets in images hidden using LSB based steganography techniques. These steganalysis tools are the DIH, RWS, and LSBMS. We shall test robustness on the same sets of stego-images, obtained at different embedding rates, from the above two image datasets.

*1) Robustness against DIH Detector*

The DIH detector estimates the length of a hidden secret in an image $I$ by measuring the correlation between the histogram of the horizontal gradient image of $I$ and the histogram of the horizontal gradient image of the image $FI$ obtained from $I$ by flipping the LSB of its pixels. It is used to detect LSB stegonography. We now report on the results of testing the same set of embedding schemes, as in the above sections, against the DIH detector using the same set of experimental cover and secret images. The charts in Figures 8 and 9 show the average DIH estimated length ratio of the embedded bit-streams at the different payload rates. These results demonstrate that the LSBM and all our mapping based embedding schemes including the Fib_IWSIM and L_IWSIM are undetectable by the DIH at all embedding rates with LSBM being the best performing scheme but only marginally better than our schemes. These experiments re-affirm the known fact that the LSBR is detectable by the DIH.

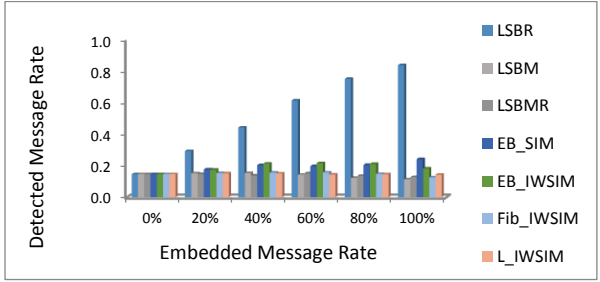

Fig. 8. Robustness against DIH - SIPI database.

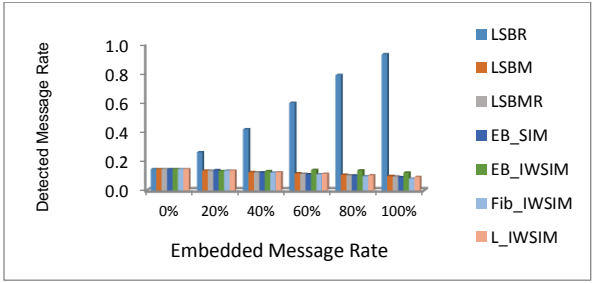

Fig. 9. Robustness against DIH - BOSSBase database.

*2) Robustness against RWS Detector*

The RWS is steganalysis tool aims to estimate the length of the secret bit-stream embedded by LSB replacement schemes. First a new image $\bar{X}$ is obtained from the suspect image $X$ is constructed. For each $\alpha \in [0, 1]$, a α-weighted stego-image is created $X^{\alpha} = (1-\alpha)X + \alpha \bar{X}$. The least square weighted approximation method is used to estimate the value of α that minimizes the Euclidian distance between $X^{\alpha}$ and the original cover image. The least square solution involves a linear filter of the input image $X$. For more detail see [16].

The charts in Figures 10 and 11 depict the average values of the estimation results of the flipped cover pixels' LSB for the tested steganography schemes for the stego SIPI database, and BOSSBase database. Similarly to the case of robustness against DIH, only the LSBR is detectable. The LSBM and all mapping based embedding including the Fib_IWSIM and L_IWSIM are undetectable by the RWS, at all embedding rates, with marginal differences between these schemes.

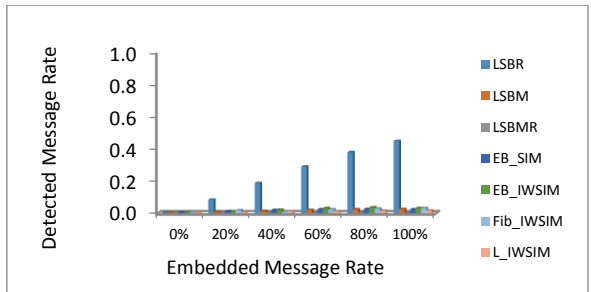

Fig.10. Robustness against RWS - SIPI database.

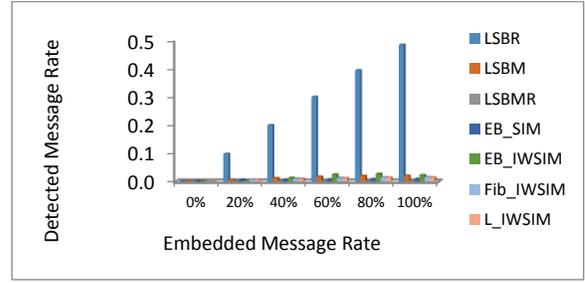

Fig. 11. Robustness against RWS - BOSSBase database.

*3) Robustness against LSBMS Detector*

The LSBMS was designed to detect the LSB matching based embedding techniques. It uses the energy distribution $H[k]$ of the histogram characteristic function (HCF), obtained from the discrete Fourier transform (DFT) of the histogram of any test image. The HCF center of mass (HCF-COM), denoted by $C(H[k])$, is calculated for k=0,…,127 by the formula:

$$(H[k]) = \frac{\sum_{i=0}^{127} i\,|H[i]|}{\sum_{i=0}^{127} |H[i]|} \quad \ldots \quad (5)$$

This tool is used to detect the hiding schemes that act as additive noise. For a grayscale test image, the ratio $C(H[k])\,/\,C(H'[k])$ of the center of mass for the image to that of a down-sampled version by a factor of two in both dimensions. When the ratio is nearly 1, the test image is declared to have no hidden secret. But if $C(H[k]) < C(H'[k])$ the image is classified as a stego.

Figures 12 and 13 show the average ratio of detected stego-images to the total number of images in the tested databases SIPI and BOSSBase, respectively.

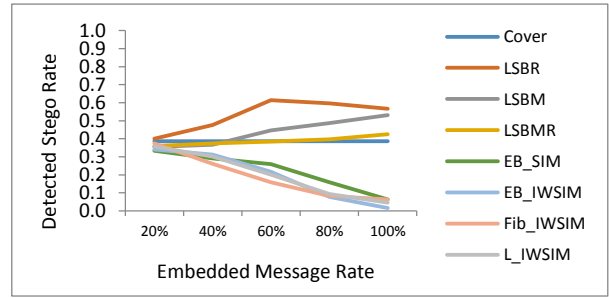

Fig. 12. Robustness against LSBMS - SIPI database.

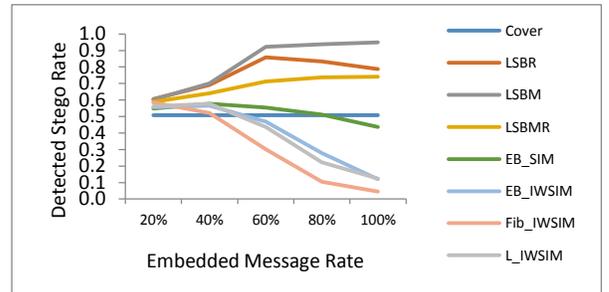

Fig. 13. Robustness against LSBMS - BOSSBase database.

Clearly, all our EB-based schemes (EB_SIM, and EB_IWSIM) as well as Fib_IWSIM, and L_IWSIM are robust against the *LSBMS* and are less detectable even than cover images at higher embedding rates. All other schemes are outperformed by our schemes, but LSBMR is best among them in that few are detected as not cover images at high embedding rate.

## V. CONCLUSIONS

In this paper, we investigated a strategy to enhance the embedding efficiency of LSB based steganography by processing both secret image and the cover image to achieve high similarity between the secret bit-stream and the cover image LSB plane. Two different secret bit-stream reversible procedures were presented that result in considerable to significant increase in ratio of 0's in the bit-stream. Several existing and new pixel decomposition schemes were shown to considerably increase the ratio of 0 LSB in cover images. Several mapping-based modification of the LSBR scheme were then proposed that exploit the expected high similarity between the secret image bit-stream and the cover image LSB plane to reduce the ratio of changed stego-image pixels. We have demonstrated that some of the mapping-based schemes achieve outperform the LSBMR scheme in terms of the steganography security measure of EE. The relatively extensive testing, demonstrated high quality stego-images and robustness against LSBR and LSBM targeted steganalysis tools.

The success of the adopted strategy that led to improve the security of LSBR by increasing embedding efficiency by about 50%, raises an interesting question as to whether the EE of other steganography schemes using such a strategy. In the future we plan to test this hypothesis by investigating ways of modifying LSBMR to improve its EE beyond what has been achieved here.

## REFERENCES

[1] I. Cox, M. Miller, J. Bloom, J. Fridrich and T. Kalker. Digital Watermarking and Steganography. 2007, Morgan Kauffman.

[2] I. J. Cox, T. Kalker, G. Pakura and M. Scheel, "Information transmission and steganography," *Proceedings of the 4th International Workshop on Digital Watermarking*, Springer, vol. 3710, pp. 15-29, 2005.

[3] D. Bhattacharyya, J. Dutta, P. Das, R. Bandyopadhyay, S. K. Bandyopadhyay and T.-H. Kim, "Discrete fourier transformation based image authentication technique," *IEEE ICCI, IEEE Computer Society*, pp. 196-200, 2009.

[4] P.-Y. Chen, H.-J. Lin and others, "A DWT based approach for image steganography," *International Journal of Applied Science and Engineering*, vol. 4, pp. 275-290, 2006.

[5] A. Westfeld, "F5—a steganographic algorithm," *International workshop on information hiding*, Springer, pp.289-302, USA, 2001.

[6] W. Luo, F. Huang and J. Huang, "Edge adaptive image steganography based on LSB matching revisited," *Information Forensics and Security, IEEE Transactions on*, vol. 5, pp. 201-214, 2010.

[7] T. Sharp, "An implementation of key-based digital signal steganography," *International Workshop on Information Hiding*, Springer, pp. 13-26, 2001.

[8] A. D. Ker, "Improved detection of LSB steganography in grayscale images," *International Workshop on Information Hiding*, Springer, pp.97-115, 2004.

[9] R. Crandall, "Some notes on steganography," *Posted on steganography mailing list*, 1998.

[10] J. Fridrich, P. Lisonek and D. Soukal, "On steganographic embedding efficiency," *International Workshop on Information Hiding*, Springer, pp 282-296, 2006.

[11] A. D. Ker, P. Bas, R. Bohme, R. Cogranne, S. Craver, T. Filler, J. Fridrich and T. Pevny, "Moving steganography and steganalysis from the laboratory into the real world," in *Proceedings of the first ACM workshop on Information hiding and multimedia security*, pp. 45-58, 2013.

[12] J. Mielikainen, "LSB matching revisited," *IEEE Signal Processing Letters*, vol. 13, pp. 285-287, 2006.

[13] C.-S. Chan, "On using LSB matching function for data hiding in pixels," *Fundamenta Informaticae*, vol. 96, pp. 49-59, 2009.

[14] M. Iranpour and F. Farokhian, "Minimal Distortion Steganography Using Well-Defined Functions," *10th International Conference on High Capacity Optical Networks and Enabling Technologies (HONET-CNS)*, pp. 21-24, Magosa, Cyprus, 2013.

[15] A. A. Abdulla, H. Sellahewa and S. A. Jassim, "Stego Quality Enhancement by Message Size Reduction and Fibonacci Bit-Plane Mapping," *International Conference on Research in Security Standardisation*, Springer, pp. 151-166, UK, 2014.

[16] A. D. Ker and R. Bohme, "Revisiting weighted stego-image steganalysis," *Proc. SPIE Electronic Imaging Security Forensics Steganography and Watermarking of Multimedia Contents*, vol. 6819, pp. 51-517. 2008.

[17] J. Fridrich and M. Goljan, "On estimation of secret message length in LSB steganography in spatial domain," *Proc. SPIE Electronic Imaging Security Forensics Steganography and Watermarking of Multimedia Contents*, pp. 23-34, 2004.

[18] T. Zhang and X. Ping, "Reliable detection of LSB steganography based on the difference image histogram," *EEE International Conference on Acoustics, Speech, and Signal Processing,. (ICASSP'03)*. vol. 1, pp. 545-548, 2003.

[19] A. D. Ker, "A general framework for structural steganalysis of LSB replacement," *International Workshop on Information Hiding*, Springer, vol. 3427, pp. 296-311, 2005.

[20] U. Viterbi, "USC-SIPI image database," *USC University of Southren California*, 1981.

[21] P. Bas, T. Filler and T. Pevny, "" Break Our Steganographic System": The Ins and Outs of Organizing BOSS," *International Workshop on Information Hiding*, Springer, vol. 6958, pp.59-70, 2011.

[22] A. Calderbank, I. Daubechies, W. Sweldens and B.-L. Yeo, "Lossless image compression using integer to integer wavelet transforms," *International Conference on Image Processing, ICIP*, pp. 596-599, 1997.

[23] D. D. L. Picione, F. Battisti, M. Carli, J. Astola and K. Egiazarian, "A Fibonacci LSB data hiding tecniqe," *14th European Signal Processing Conference,EUSIPCO*, pp. 1-5, 2006.

[24] S. Dey, A. Abraham and S. Sanyal, "An LSB Data Hiding Technique Using Prime Numbers," *Third International Symposium on Information Assurance and Security, IAS 2007, IEEE*, pp. 101-108, 2007.

[25] S. Dey, A. Abraham and S. Sanyal, "An LSB Data Hiding Technique Using Natural Number Decomposition," *International Conference on,Intelligent Information Hiding and Multimedia Signal Processing,. IIHMSP 2007. IEEE*, vol. 2, pp. 473-476, 2007.

[26] F. Alharbi, "Novel Steganography System using Lucas Sequence," *International Journal of Advanced Computer Science and Applications (IJACSA)*, vol. 4, pp. 52-58, 2013.

[27] N. Aroukatos, K. Manes, S. Zimeras and F. Georgiakodis, "Data hiding techniques in steganography using fibonacci and catalan numbers," 9[th] *International Conference on Information Technology: New Generations (ITNG),IEEE*, pp. 392-396, 2012.

[28] A. A. Abdulla, H. Sellahewa and S. A. Jassim, "Steganography based on pixel intensity value decomposition," *Proc. SPIE Electronic Imaging Security Forensics Steganography and Watermarking of Multimedia Contents*, pp. 912005-912005, 2014.

[29] A. A. Abdulla, S. A. Jassim and H. Sellahewa, "Efficient high-capacity steganography technique," *Proc. SPIE Electronic Imaging Security Forensics Steganography and Watermarking of Multimedia Contents*, pp., 875508-875508, 2013.

[30] A. A. Abdulla. "Exploiting Similarities between Secret and Cover Images for Improved Embedding Efficiency and Security in Digital Steganography," PhD dissertation, Dept. of Applied Computing, Buckingham Univ., Buckingham, UK, 2015.